\documentclass[galaxies,article,accept,moreauthors]{Definitions/mdpi}

\graphicspath{{Definitions/}}

\newcommand{\newathena}{\mbox{\em NewAthena\/}}
\usepackage{xcolor}

\newcommand{\chandra}{\mbox{\em Chandra\/}}
\newcommand{\spinpar}{$a^*$}
\newcommand{\xrism}{\mbox{\em XRISM\/}}
\newcommand{\xmm}{\mbox{\em XMM-Newton\/}}
\newcommand{\nustar}{\mbox{\em NuSTAR\/}}
\newcommand{\suzaku}{\mbox{\em Suzaku\/}}
\newcommand{\hexp}{\mbox{\em HEX-P\/}}
\newcommand{\swift}{\mbox{\em Swift\/}}
\newcommand{\nuwelldefined}{10}
\newcommand{\nulowerbounds}{28}
\newcommand{\numallick}{13}

\newcommand{\nufullsample}{51}

\usepackage{amssymb}
\firstpage{1} 
\pubvolume{1}
\issuenum{1}
\articlenumber{0}
\pubyear{2026}
\copyrightyear{2026}
\externaleditor{ } %
\datereceived{ } 
\daterevised{ } 
\dateaccepted{ } 
\datepublished{ } 
\Title{Spin Demographics of Active Supermassive Black Holes: Updated Estimates from X‑Ray Reflection and Future~Opportunities}

\Author{%
J\'ulia M. Sisk-Reyn\'es $^{1,}$*\orcidA{}, 
Christopher S. Reynolds $^{2}$\orcidB{}, 
James H. Matthews $^{3}$\orcidC{}, 
Dominic J. Walton $^{4}$\orcidD{}, 
\mbox{Joanna M. Piotrowska $^{5}$\orcidE{}}, 
James F. Steiner $^{1}$\orcidF{}, 
Javier A. Garc\'ia $^{6}$\orcidG{} and 
Angelo Ricarte $^{1,7}$\orcidH{}}

\AuthorNames{J\'ulia Sisk-Reyn\'es et al.}

\address{%
$^{1}$ \quad Center for Astrophysics | Harvard \& Smithsonian, Cambridge, MA 02138, USA; james.steiner@cfa.harvard.edu~(J.F.S.); angelo.ricarte@cfa.harvard.edu~(A.R.)\\
 
$^{2}$ \quad Department of Astronomy, University of Maryland, College Park, MD 20742, USA; creynold@umd.edu\\

$^{3}$ \quad Department of Physics, Astrophysics, University of Oxford, Denys Wilkinson Building, Keble Road, \mbox{Oxford OX1 3RH, UK}; james.matthews@physics.ox.ac.uk\\

$^{4}$ \quad Centre for Astrophysics Research, Department of Physics, Astronomy and Mathematics, University of Hertfordshire, College Lane, Hatfield AL10 9AB,
UK; d.walton4@herts.ac.uk\\

$^{5}$ \quad Cahill Center for Astronomy \& Astrophysics, California Institute of Technology, Pasadena, CA 91125, USA; jpiotrowska.astro@gmail.com\\

$^{6}$ \quad X-Ray Astrophysics Laboratory, NASA Goddard Space Flight Center, Greenbelt, MD 20771, USA; javier.a.garciamartinez@nasa.gov\\

$^{7}$ \quad Black Hole Initiative at Harvard University, 20 Garden Street, Cambridge, MA 02138, USA\\
}

\corres{Correspondence: julia.sisk\_reynes@cfa.harvard.edu}

\abstract{Understanding the growth of supermassive black holes (SMBHs) requires observational constraints on how their angular momentum---or spin---varies with mass, since the relative importance of coherent accretion, chaotic accretion, and mergers will be reflected in SMBH spin populations. Here we present an updated compilation of reflection‑based SMBH spin measurements from the literature and assemble a set of ancillary quantities of interest for each SMBH (including redshift, Eddington ratio, and X‑ray luminosity). No obvious apparent correlation between the Eddington-scaled accretion rate and the black hole spin is seen, noting that formal statistical tests are beyond the scope of this review. We discuss the limitations of using this heterogeneous mass--spin sample to test predictions of SMBH growth from semi‑analytic models and cosmological simulations, emphasizing the need for a more uniform sample. We then highlight the encouraging prospects enabled by the next-generation \newathena\ X-ray flagship observatory. Finally, we summarize how hierarchical Bayesian population inference applied to observed SMBH mass--spin populations will constitute a powerful framework for confirming tentative mass--spin trends in future samples.}

\begin{document}

\section{Introduction}
\label{sec1}

The no-hair theorem of General Relativity states that astrophysical (uncharged) black holes are described by two fundamental quantities: black hole mass, $M_\mathrm{BH}$, and angular momentum, $\textbf{J}$, or spin. The spin is commonly expressed as a dimensionless parameter {\spinpar}~=~$c~\textbf{J}/G_\mathrm{N}~M_\mathrm{BH}^{2}$, where $c$ is the speed of light in the vacuum and $G_\mathrm{N}$ is Newton's gravitational constant. For a Kerr (spinning) black hole, \spinpar\ must be within $\pm 1$, where negative (positive) values of \spinpar\ denote orbits {of the accretion disk} that are counter-rotating (co-rotating) with respect to the black hole spin. {Early theoretical work established two distinct pathways for black hole spin evolution.~As first demonstrated by Ref.~\cite{bardeen-1973}, a black hole is spun up by prolonged prograde coherent accretion of the incoming material adding angular momentum in the same direction. By contrast, Refs.~\cite{press-1972, hawking-1972}
had demonstrated that a black hole will be spun down by the capture of counter-rotating gas and photon orbits or chaotic accretion of matter falling in from different directions. Ref.~\cite{thorne-1974} showed that a hole spun up by prolonged prograde coherent accretion will inevitably be spun down by the capture of counter-rotating gas and photon orbits, imposing the canonical upper limit on the spin magnitude of a Kerr black hole of \spinpar $=+0.998$ accreting from a{\linebreak} Novikov-Thorne disk.}

\textls[-8]{In addition to being a fundamental property, the spins of supermassive black holes (SMBHs) in active galactic nuclei (AGN) act as fossil records of SMBH growth. {Since the mass alone does not allow distinguishing between different cosmological black hole growth processes (and precisely because different processes, e.g., mergers or gas accretion are expected to yield different spin distributions across black hole mass), the spin demographics of SMBH populations are unique observational probes of their growth histories~\cite{reynolds-spinreview}.}~In some semi-analytic models (SAMs) of hierarchical structure formation, SMBHs with $\mathrm{log}(M_\mathrm{BH}/M_\odot) \lesssim 8$ primarily grow via {coherent} accretion via {thin disks}, leading to high-spin SMBHs in galaxies like the Milky Way~\cite{volonteri-2005,sesana-2014}. {Here, `coherent' accretion refers to infalling matter whose orbits are aligned with the spin axis of the black hole. In contrast, }other SAMs also show that a prolonged phase of {chaotic gas accretion would spin black holes down~\cite{dotti-2013} on timescales of $\sim${few} {Gyr}---where `chaotic' refers to a scenario where material falls onto the black hole following orbits at a different orientation to the black hole spin axis}. For instance, Ref.~\cite{king-2008} showed that late-time {chaotic} accretion will lower the average spin of SMBHs in local AGN across mass scales. Since the characteristic spin-up and spin-down timescales in SAMs are frequently sensitive to the physical (rather than Eddington) accretion rate $\dot{m}$, transitions between radiatively inefficient and efficient accretion modes can further accelerate or suppress this mass--spin evolution~\cite{bustamante-2019,beckmann-2025}.}

{Spinning SMBHs threaded by a magnetic field can also lose angular momentum through the jet. As derived by Ref.~\cite{blandford-znajek-1977}, the Blandford-Znajek (BZ) jet efficiency scales with both the square of the spin {magnitude} and the square of the magnetic field strength, $\eta\propto (a^*)^2B^2$, with potential corrections at large spin magnitudes~\cite{tchekhovskoy-2010,tchekhovskoy-2011}.}
~Recently, by fitting General Relativistic Magnetohydrodynamics (GRMHD) simulations, Ref.~\cite{ricarte-2023} proposed a model in which significant spin-down occurs whenever the disk becomes geometrically thick, at either highly sub-Eddington or super-Eddington accretion rates{, building upon the formalism of Ref.~\cite{lowell-2024} with the inclusion of radiative simulations in the super-Eddington regime. The resultant formulae for both jet efficiency and spin evolution were then self-consistently placed in a SAM,} demonstrating observationally testable spin moderation via this process even when accretion proceeds in a purely coherent fashion~\cite{ricarte-2025}. Exploring the magnitude of BZ-driven spin-down in different simulation setups is an active area of research. Using GRMHD simulations with radiative cooling, Ref.~\cite{lowell-2025} proposed a {low} universal equilibrium spin value of $a^*\approx 0.3$ for luminous strongly magnetized accretion flows. Meanwhile, using multi-zone GRMHD simulations from the event horizon to the Bondi radius, Ref.~\cite{cho-2026} {found} less constant jets, and therefore longer equilibrium timescales than Refs.~\cite{narayan-2022,ricarte-2023} by a factor of a few.

Over the past decade, hydrodynamical simulations of cosmic structure formation have started incorporating sub-grid prescriptions to account for SMBH spin evolution. Recently, Ref.~\cite{sala-2024} ran a cosmological simulation with \textsc{OpenGadget3} code using a novel sub-resolution prescription to track the black hole spin by accounting for the effects of coalescence and misaligned accretion through a geometrically thin, optically thick accretion disk. Ref.~\cite{sala-2024} found that low-mass holes ($M_\mathrm{BH} < 10^7\,M_\odot$) grow primarily through gas accretion, occurring mostly in a coherent fashion that favors spin-up. At higher masses ($M_\mathrm{BH} > 10^7\,M_\odot$), the gas angular momentum directions of subsequent accretion episodes were often found to be uncorrelated. A high level of correlation between counter-rotating accretion and black hole spin-down was thus inferred at masses >10$^7\,M_\odot$---a regime where SMBH coalescence was also identified to be an important growth channel. Overall, Ref.~\cite{sala-2024} concluded that the spin distributions from their simulation display a wide variety of histories, depending on the dynamical state of the gas feeding the black hole and the relative contribution of mergers and gas accretion. Other state-of-the-art numerical models also highlight that the efficiency of spin evolution is strongly tied to the instantaneous accretion rate: at fixed $\dot{m}$, low-mass SMBHs can undergo rapid spin-up on short timescales, whereas massive SMBHs require substantially longer periods of sustained coherent inflow to appreciably change their spin~\cite{beckmann-2025}.

Here we outline the prospects of utilizing observed SMBH mass--spin populations with current and future samples to test predictions of SMBH growth from SAMs and hydrodynamic simulations, where {spin estimates} are drawn from X-ray reflection spectroscopy. These reflection-based estimates are expected to trace the innermost flow onto holes whose surrounding accretion disks are geometrically thin and optically thick in the Shakura-Sunyaev regime~\cite{shakura-sunyaev,novikov-thorne}. Therefore, we do not consider spin estimates based on other methods \endnote{Other spin inference methods may include VLBI imaging and polarization signatures, using empirical and fundamental-plane relations in AGN samples, using a thick disk interpretation to describe the soft X-ray spectrum of tidal disruption events; and SED fitting (see Refs.~\cite{daly-2016,unal-2020,daly-2022,cao-2023,temple-2023,palumbo-2025}).}. We have made the set of archival SMBH mass and reflection-inferred spin estimates compiled here publicly available at the Github repository \href{https://github.com/joanna-pk/xray-reflection-spin-repository}{https://github.com/joanna-pk/xray-reflection-spin-repository} (accessed on 10 May 2026) to enable continuous updates as new constraints become available or existing ones are revisited. {Since the current sample is heterogeneous and affected by large, heteroscedastic uncertainties in spin, here do not attempt formal correlation tests. A full statistical treatment (including forward modeling and population‑level inference) will be presented in a separate contribution within a \newathena\ Special Issue currently under preparation for publication in \textit{JHEAp} by{\linebreak} 2027~\cite{newathena-inprep}}.

This review is organized as follows. In Section~\ref{sec2}, we present the updated SMBH mass--spin sample with reflection-inferred spins compiled from the literature, together with several ancillary quantities of interest---including redshift, Eddington ratio and X-ray luminosity (2--10~keV observed frame). We then interpret the mass--spin plane and find no obvious correlation between the Eddington-scaled accretion rate and the black hole spin. We outline the caveats associated with using the current mass--spin sample to test predictions of SMBH growth models which predict that accretion-driven and accretion+merger-driven growth would imprint different expected trends in the {mass--spin} plane. The caveats we highlight arise from the present limitations in sample size, heterogeneity, and statistical and systematic uncertainties. In Section~\ref{sec3}, considering the heterogeneity of reflection-based spin inference in the current literature, we argue that hierarchical Bayesian inference approaches hold a promising pathway to confirming the presence of possible mass--spin trends in current and future observed populations. Finally, we introduce \newathena's~encouraging prospects in enabling such an assessment. {We conclude in Section~\ref{sec4}.}

\newathena\ is the European Space Agency’s next‑generation flagship‑class X‑ray observatory, planned for launch in 2037 with unprecedented survey and spectroscopic capabilities~\cite{newathena-5, newathena-1, newathena-3, newathena-2, newathena-4}. With its large collecting area, broad bandpass (0.1--12~keV), and the exceptional spectral resolution of its X‑IFU microcalorimeter ({$\leq$4~eV at 7~keV}), \newathena\ will resolve broad and narrow reflection features in the iron K band of nearby AGN that could be degenerate with relativistically smeared reflection. {The two instruments aboard \newathena---WFI and X-IFU---will have reference collecting areas of $9300 \ {\mathrm{cm}}^{2}$ and $5900 \ {\mathrm{cm}}^{2}$ respectively---jointly representing a factor of >2 improvement over the joint collecting area of the three instruments aboard \xmm, and over a factor of 100 on that of \chandra/Gratings.} \newathena\ will also extend reflection-based spin inference into a high redshift regime ($ z \lesssim 1.5$) in distant AGN whose Fe K band is redshifted into \newathena's bandpass. \newathena's {anticipated} strategic survey of at least 50 new nearby SMBHs is expected to deliver the pathway towards a robust observational discrimination between accretion-driven versus accretion+merger-driven growth from observed mass--spin trends in the local universe.

\section{The Observed SMBH Mass vs. Spin Plane}
\label{sec2}

\begin{figure}[H]
\includegraphics[width=13.8cm]{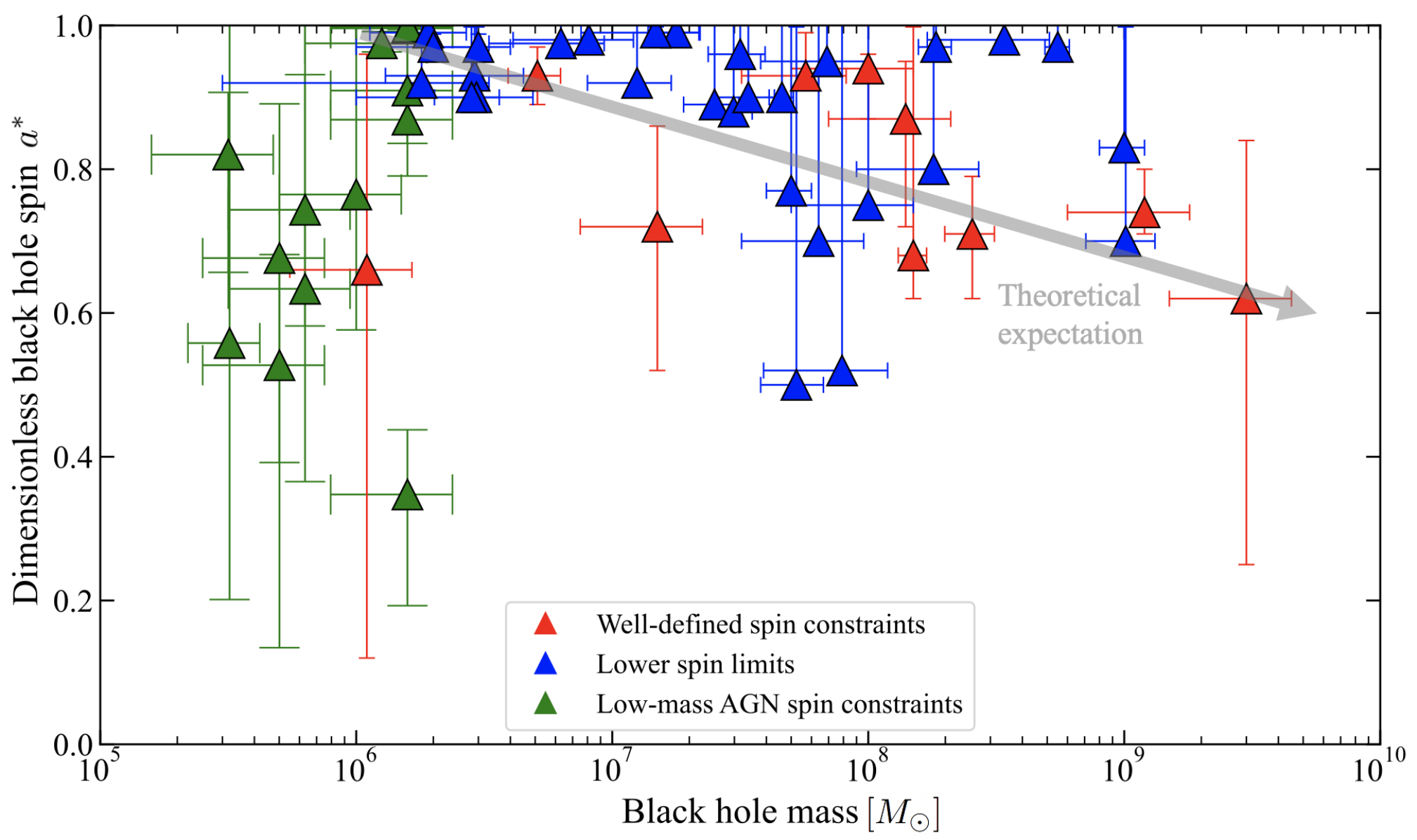}
\caption{Observed SMBH mass--spin plane with reflection-inferred spins compiled from published literature (data listed in Table~\ref{tab1}; error bars in spin and mass show the 90\% and 68\% confidence levels, respectively). The plane comprises: \nuwelldefined\ (red) and \nulowerbounds\ (blue) SMBHs with well-defined vs. lower spin bounds updated from the spin reviews in Refs.~\cite{bambi-spinreview,reynolds-spinreview}; and \numallick\ low-mass AGN (green) presented in Ref.~\cite{mallick-2022} {(where 12 have well-defined spins)}. The gray arrow marks the expectation from theory.\label{fig1}}
\end{figure}

{The spin vs. mass plane for most moderately accreting SMBHs with existing spin estimates from X‑ray reflection spectroscopy---compiled from published or accepted journal articles at the time of writing---is shown in Figure~\ref{fig1}. This distribution appears to show a tentative decrease in spin at masses $\geq$10$^8\,M_\odot$, although the current sample size and selection effects mean this should be interpreted with caution. Such a pattern could be consistent with expectations from merger‑driven growth at these higher masses, but the data do not yet allow a firm conclusion. In contrast, a distinct low‑mass ($10^{6\text{--}7}\,M_\odot$) population of SMBHs with high‑to‑maximal spins ({\spinpar} $\sim 0.998$) may be emerging, potentially suggestive of coherent‑accretion‑driven growth~\cite{dubois-2021,beckmann-2025}. This possible trend across black hole mass---oriented by theoretical predictions---is illustrated by the gray arrow in Figure~\ref{fig1}. If not solely the result of selection effects, the absence of retrograde spins in the current sample may tentatively disfavor the contribution of prolonged chaotic accretion, since this would be expected to yield a broader mass–spin distribution (including retrograde values). The presence of SMBHs with high‑to‑maximal spins in the current sample is broadly consistent with expectations from coherent accretion scenarios, though the current data remain insufficient to draw statistically robust conclusions. By contrast, the exclusion of maximal spin values at 90\% confidence in several SMBHs could be qualitatively suggestive that these SMBHs are spun down by other processes, e.g., mergers}.

\vspace{-1pt}
\vspace{-8pt}

\begin{table}[H] 
\caption{Spin
 and mass estimates for 51 SMBHs from an updated literature compilation of {Ref.~\cite{reynolds-spinreview,bambi-spinreview}} (where the first 22 rows distinguish those with well-defined spin estimates). SMBHs whose name is accompanied by an asterisk were taken from the low-mass sample of AGN studied in Ref.~\cite{mallick-2022} (see Table 1 of Ref.~\cite{mallick-2022} for each source's full name). Rows are ordered by decreasing mass. The error bars in mass and spin correspond to the 68\% and 90\% statistical uncertainties, respectively. The third column in the table references the respective papers from which we quote the mass and spin estimates for each source (mass and spin estimates were taken from a single paper if a single citation appears). The last column indicates the optical spectral classification, where we make use of the following abbreviations: Rq: radio-quiet; Ri: radio-intermediate; Q: quasar; 1 (2): type-1 (type-2); Sy: Seyfert; BLRG: broad-line radio galaxy; NL: narrow-line; BL: broad-line; BAL: broad-absorption-line; Lensed: gravitationally lensed.}

\label{tab1}
\small
\begin{tabularx}{\linewidth}{%
m{3cm}<{\centering}
m{3.1cm}<{\centering}
m{1.9cm}<{\centering}
m{1.8cm}<{\centering}
m{2cm}<{\centering}
}
\toprule
\textbf{Source} & \boldmath{$M_\mathrm{BH}[10^6M_\odot]$} & \boldmath{\textbf{Spin $a^*$}} & \textbf{Refs.} & \textbf{Type}\\
\midrule

H 1821+643~$^{\dagger\dagger}$ & $(3.0^{+1.5}_{-1.5})\times 10^3$ & $0.62^{+0.22}_{-0.37}$ & \cite{shapovalova-2016,sisk-reynes-2022} & RqQ,1  \\
Q 2237+305~$^{\dagger\dagger}$ & $(1.2^{+0.6}_{-0.6})\times 10^3$ & $0.76^{+0.06}_{-0.03}$ & \cite{hutsemekers-sluse-2021, mark-reynolds-2014} & Lensed Q,1\\
Fairall 9~$^{\dagger}$ & $255^{+56}_{-56}$ & $0.71^{+0.08}_{-0.09}$ & \cite{peterson-2004-sample,lohfink-2012} & NL,Sy,1\\
Ark 120~$^{\dagger}$ & $150^{+19}_{-19}$ & $0.64^{+0.32}_{-0.06}$ & \cite{peterson-2004-sample,porquet-2019} & Sy1\\
RX J1131-1231~$^{\dagger}$ & $140 \pm 70$ & $0.87^{+0.08}_{-0.15}$ & {\cite{reis-2014}} & Lensed Q,1\\
IRAS 09149–6206 * & $(1.0^{+0.5}_{-0.5})\times 10^2$ & $0.94^{+0.02}_{-0.07}$ & \cite{vasudevan-2016-sample,walton-2019} & Sy1\\
PG 1229+204~$^{\dagger}$ & $57^{+25}_{-25}$ & $0.93^{+0.06}_{-0.02}$ & \cite{peterson-2004-sample,jiang-2019-spinsample} & Sy1 \\ 
Swift J2127.4+5654 & $15.0^{+7.5}_{-7.5}$ & $0.72^{+0.14}_{-0.20}$ & \cite{vasudevan-2016-sample,jiang-2019-spinsample} & Sy1 \\ 
NGC 5506 & $5.1^{+1.18}_{-1.18}$ & $0.93^{+0.04}_{-0.04}$ & \cite{nikolajuk-2009,sun-2018} & NL,Sy1 \\ 
Mrk 359~' & $1.10^{+0.55}_{-0.55}$ & $0.66^{+0.30}_{-0.54}$ & \cite{zw-2005-sample,walton-2013-spinsample} & NL,Sy1 \\
J0107~$^{\dagger\dagger}${(*)} & $10^{-1} \times (16.0^{+16.0}_{-8.0})$ & ${0.87}^{+0.08}_{-0.24}$ & \cite{greene-2007-sample,mallick-2022} & NL,Sy1 \\
J0940~$^{\dagger\dagger}$(*) & $10^{-1} \times (16.0^{+16.0}_{-8.0})$ & ${0.996}^{+0.001}_{-0.015}$ & \cite{greene-2007-sample,mallick-2022} & NL,Sy1 \\ 
J1357~$^{\dagger\dagger}$(*) & $10^{-1} \times (16.0^{+16.0}_{-8.0})$ & ${0.35}^{+0.15}_{-0.09}$ & \cite{greene-2007-sample,mallick-2022} & NL,Sy1 \\ 
J1541~$^{\dagger\dagger}$(*) & $10^{-1} \times (16.0^{+16.0}_{-8.0})$ & ${0.91}^{+0.07}_{-0.21}$ & \cite{greene-2007-sample,mallick-2022} & BL,Sy1  \\ 
J1140~$^{\dagger\dagger}$(*) & $10^{-1} \times (12.6^{+12.5}_{-6.3})$ & ${0.975}^{+0.012}_{-0.016}$ & \cite{greene-2007-sample,mallick-2022} & NL,Sy1  \\ 
J1347~$^{\dagger\dagger}$(*) & $10^{-1} \times (10.0^{+10.0}_{-5.0})$ & ${0.77}^{+0.19}_{-0.43}$ & \cite{greene-2007-sample,mallick-2022} & NL,Sy1  \\ 
J1434~$^{\dagger\dagger}$(*) & $10^{-1} \times (6.3^{+6.3}_{-3.1})$ & ${0.63}^{+0.27}_{-0.45}$ & \cite{greene-2007-sample,mallick-2022} & Sy1  \\
J1631~$^{\dagger\dagger}$(*) & $10^{-1} \times (6.3^{+6.3}_{-3.1})$ & ${0.76}^{+0.16}_{-0.19}$ & \cite{greene-2007-sample,mallick-2022} & BL,Sy1  \\ 
J1023~$^{\dagger\dagger}$(*) & $10^{-1} \times (5.0^{+5.0}_{-2.5})$ & ${0.53}^{+0.39}_{-0.15}$ & \cite{greene-2007-sample,mallick-2022} & NL,Sy1  \\ 
J1626~$^{\dagger\dagger}$(*) & $10^{-1} \times (5.0^{+5.0}_{-2.5})$ & ${0.68}^{+0.28}_{-0.21}$ & \cite{greene-2007-sample,mallick-2022} & Sy1.5  \\ 
J0228~$^{\dagger\dagger}$(*) & $10^{-1} \times (3.2^{+3.1}_{-1.6})$ & ${0.82}^{+0.16}_{-0.09}$ & \cite{greene-2007-sample,mallick-2022} & BL,Sy1  \\ 
POX 52~$^{\dagger\dagger}$(*) & $10^{-1} \times (3.2^{+3.1}_{-1.6})$ & ${0.56}^{+0.36}_{-0.46}$ & \cite{thornton-2008,mallick-2022} & Sy1.8  \\ \midrule
PG 1426+015~$^{\dagger}$ & $(1.0^{+0.3}_{-0.3})\times 10^3$ & >0.70 & \cite{peterson-2004-sample,walton-2025} & Rq,Sy1\\ 
PG 2112+059 & $(1.0^{+0.2}_{-0.2})\times 10^3$ & >0.83 & \cite{vestegaard-2006-sample,schartel-2010} & BAL,Q\\ 
PG 0804+761~' & $550^{+60}_{-60}$ & >0.97 & \cite{zw-2005-sample,jiang-2019-spinsample} & Rq,1\\
{1 H0419–577} & $340^{+170}_{-170}$ & >0.98 & \cite{grupe-2010,jiang-2018-spins-1h0419} & Rq,Sy1 \\
Mrk 1501~$^{\dagger\dagger}$ & $184^{+27}_{-27}$ & >0.97 & \cite{grier-2012-sample,chamani-2020} & Ri,1\\
RBS 1124~$^{\dagger}$ & $180^{+90}_{-90}$ & >0.236 & \cite{peterson-2004-sample,madathil-pottayil-2024} & Rq,Q\\
Fairall 51 & $(1.0^{+0.5}_{-0.5})\times 10^2$ & >0.6 & \cite{bennert-2006-sample,svoboda-2015} & Sy1\\
Mrk 841 & $79^{+40}_{-40}$ & >0.52 & \cite{zw-2005-sample,walton-2013-spinsample} & Rq,Sy1 \\
IRAS 13197-1627~' & $64^{+34}_{-34}$ & >0.7 & \cite{zw-2005-sample,walton-2018} & Sy1.8\\
3C~120~$^{\dagger}$ &$60^{+31}_{-31}$ & >0.95 & \cite{peterson-2004-sample,lohfink-2013} & BLRG \\ 
Mrk 79~$^{\dagger}$ & $52.4^{+14.4}_{-14.4}$ & >0.5 & \cite{peterson-2004-sample,jiang-2019-spinsample} & Sy1.2 \\ 
{IRAS 0521–7054~'} & $50^{+10}_{-10}$ & >0.77 & {\cite{walton-2020}} & Sy2 \\
NGC 4151~$^{\dagger}$ & $50^{+10}_{-10}$ & >0.9 & \cite{keck-2015,bentz-2022} & Sy1.5 \\ 
1 H0323+342~$^{\dagger}$ & $34^{+9}_{-9}$ & >0.9 & \cite{wang-2016,ghosh-2018} & NL,Sy1\\
ESO 033-G002~' & $31.6^{+7.9}_{-7.9}$ & >0.96 & {\cite{walton-2021}} & Rq,Sy2\\
NGC 3783~$^{\dagger\dagger\dagger}$ & $25.4^{+9.0}_{-7.2}$ & >0.88 & \cite{amorim-2021,brenneman-2011} & BAL,Sy1 \\ 
Mrk 110~$^{\dagger}$ & $25.1^{+6.1}_{-6.1}$ & >0.89 & \cite{peterson-2004-sample,jiang-2019-spinsample} & NL,Sy1 \\
Mrk 335~$^{\dagger}$ & $17.8^{+4.0}_{-4.0}$ & >0.91 & \cite{peterson-2004-sample,gallo-2015} & NL,Sy1 \\
PG 1535+547 & $14.8^{+7.2}_{-7.2}$ & >0.99 & \cite{hu-2021-sample, madathil-pottayil-2026} & NL,Sy1 \\ 
{ESO 362–G18} & $12.5^{+4.5}_{-4.5}$ & >0.92 & {\cite{agis-gonzalez-2014}} & Sy1.5\\
Tons 180~' & $8.1^{+4.0}_{-4.0}$ & >0.98 & \cite{zw-2005-sample,jiang-2019-spinsample} & NL,Sy1\\
{IRAS 13224–3809~'} & $6.3^{+3.0}_{-3.0}$ & >0.975 & \cite{zw-2005-sample,jiang-2018-spins-iras13224} & NL,Sy1\\
1 H0707-495~' & $3.0^{+1.0}_{-1.0}$ & >0.97 & \cite{zw-2005-sample,zoghbi-2010} & NL,Sy1\\

\bottomrule
\end{tabularx}
\end{table}

\begin{table}[H] \ContinuedFloat
\caption{{\em Cont.}} \label{tab1}
\small
\begin{tabularx}{\linewidth}{%
m{3cm}<{\centering}
m{3.1cm}<{\centering}
m{1.9cm}<{\centering}
m{1.8cm}<{\centering}
m{2cm}<{\centering}
}
\toprule
\textbf{Source} & \boldmath{$M_\mathrm{BH}[10^6M_\odot]$} & \boldmath{\textbf{Spin $a^*$}} & \textbf{Refs.} & \textbf{Type}\\
\midrule

{MCG–06-30-15}~$^{\dagger}$ & $2.9^{+1.6}_{-1.6}$ & >0.65 & \cite{bentz-2016,brenneman-2025} & NL,Sy1\\
Mrk 1044 & $2.82^{+0.90}_{-0.73}$ & >0.9 & \cite{du-2015,mallick-2018} & NL,Sy1
\\ 
Ark 564 & $2.3^{+2.6}_{-1.3}$ & >0.9 & \cite{lewin-2022, jiang-2019-spinsample} & NL,Sy1\\
NGC 1365 & $2.0^{+1.0}_{-1.0}$ & >0.97 & \cite{risaliti-2009, walton-2014} & Sy1.5--1.8\\
Mrk 766~' & $1.8^{+0.5}_{-0.5}$ & >0.92 & \cite{zw-2005-sample,buisson-2018} & NL,Sy1\\

J1559~$^{\dagger\dagger}$(*) & $10^{-1} \times (16.0^{+16.0}_{-8.0})$ & >0.975 & \cite{greene-2007-sample,mallick-2022} & NL,Sy1  \\ 

\bottomrule
\end{tabularx}
\noindent{\footnotesize{\textls[-15]{
 Black hole mass estimated via:~optical reverberation mapping~$^\dagger$; $H\alpha$, $H\beta$ or $C\texttt{IV}$ widths, combined with a subsequent fit of virial scaling relations~$^{\dagger\dagger}$; VLTI GRAVITY interferometry~$^{\dagger\dagger\dagger}$; an empirical method based on observed correlations between the equivalent width attributed to narrow-line 6.4 Fe K$\alpha$ {emission} '; and other methods (no symbol).}}}
\end{table}

\subsection{Updated Mass--Spin Plane}

The mass--spin estimates shown in Table~\ref{tab1} are an updated version of Table 1 of the spin review of Ref.~\cite{reynolds-spinreview}, incorporating the following changes: 

\begin{itemize}
\item For the high-mass SMBH H 1821+643, we adopt the spin estimate in Ref.~\cite{sisk-reynes-2022} (consistent with the prior bound of Ref.~\cite{reynolds-2014-1821}). We note that Ref.~\cite{yaqoob-2005} argued that the iron K band can be described with a model featuring absorption and distant reflection.
\item For the extreme galaxy ESO 033–G002, we quote the mass and spin reported in Ref.~\cite{walton-2021}---consistent with the spin later estimated by Ref.~\cite{nekrasov-2025} under a disk reflection spectrum for an extended (ring) coronal geometry.
\item For Fairall 9, we consider the spin estimate inferred from spectral modeling of multi-epoch \xmm\ and \suzaku\ observations of Ref.~\cite{lohfink-2012} without the inclusion of a model component for the soft excess in the \suzaku\ data (as such an inclusion otherwise drives the spin constraint, as detailed in their discussion). We note that several works have argued that relativistically-broadened Fe K$\alpha$ emission is not required to describe the X-ray spectrum~\cite{yaqoob-2016} or the X-ray variability~\cite{hagen-2023} of Fairall 9.
\item For the Seyfert~1.5 galaxy NGC 4151, we adopt the lower spin bound {\spinpar} > 0.9 found from an X-ray reflection fit to a joint \swift+\suzaku\ spectrum which assumed a lamppost coronal geometry~\cite{keck-2015}. Whilst this geometry seems to be strongly disfavored by joint \textit{IXPE}, \xmm, and \nustar\ polarimetric and spectroscopic analyses~\cite{gianolli-2023,gianolli-2024}, a 2023 \xrism\ observation does reveal relativistically broadened Fe K$\alpha$ emission. A new spin constraint from this \xrism\ observation is anticipated \citep{xrism-4151-2024}.
\item We include 13 low-mass AGN sample spin estimates in Ref.~\cite{mallick-2022}, who used a high-density disk reflection model to describe the soft excess in \xmm\ data. 
\item We do not include the spin constraints for both IRAS~13349+2438 and the high-mass broad-line radio galaxy 4C~74.26 for the reasons outlined in Section 6 of Ref.~\cite{sisk-reynes-2022}.
\item We do not consider the spin estimate for NGC 4051 of Ref.~\cite{patrick-2012}, as its spin was fixed to the canonical upper limit in their spectral analysis.
\item {For the canonical type-1 AGN MCG–6-30-15, we adopt the recent time-resolved spin estimate of Ref.~\cite{wilkins-2026} ({\spinpar} > 0.93) from a quasi-simultaneous \xrism, \xmm, and \nustar\ campaign. We note this value is consistent with the lower spin bound obtained from a time-averaged analysis of these data~\cite{brenneman-2025}. We note that work prior to the launch of \xrism\ had also inferred high-spin lower bounds for this AGN~\cite{brenneman-2006,marinucci-2014-b}.}
\item We update the mass--spin compilation of Ref.~\cite{reynolds-spinreview} with four new SMBHs: Mrk 1044~\cite{mallick-2018}, {ESO 033–G002}~\cite{walton-2021}, PG 1426+015~\cite{walton-2025}, PG 1535+547~\cite{madathil-pottayil-2026}.
\end{itemize}

\subsection{Interpretation of the Observed Mass--Spin Plane}
The large statistical uncertainties of many existing spin estimates make a robust assessment of possible trends challenging. High-mass SMBH spin measurements---where merger-driven spin-down is expected to be most apparent---also remain scarce. Within the full sample of \nufullsample\ accreting SMBHs and low-mass AGN, only 22 have well-defined upper and lower bounds, while the remaining 29 only have lower limits.

Most existing reflection-based SMBH spin measurements have been obtained on a case-by-case basis, resulting in a heterogeneous dataset with varying reflection-model assumptions and implementations. The majority of the spin estimates in Figure~\ref{fig1} were obtained using variants of the \textsc{Relxill} relativistic X-ray reflection model \citep{dauser-2014,garcia-2014,dauser-2016,dauser-2022}, applied to either broadened Fe K$\alpha$ emission and the Compton hump, or to the soft excess. Close to half of all existing spin estimates (excluding the low-mass sample from Ref.~\cite{mallick-2022}) were based on broadband X-ray spectra covering both the Fe K band and the Compton hump regimes. In only approximately half of these cases (i.e., $\sim$25\% of all 51 sources in the full sample), quasi-simultaneous \xmm+\nustar\ observations provide the most comprehensive data: \xmm's spectral sensitivity is critical to probing the red wing of the Fe K line, while \nustar's high-energy coverage (whose bandpass covers 3--79~keV) constrains the Compton hump, which is essential for reducing spectral modeling degeneracies.

At present, the disk reflection spectrum of only one accreting SMBH ({MCG–06-30-15}) has been probed with joint \xrism+\xmm+\nustar\ coverage~\cite{brenneman-2025}, although similar sensitivity is expected for ongoing and future \xrism\ targets (including NGC 4151; see Ref.~\cite{xrism-4151-2024}). With the fine spectral resolution of \xrism/Resolve ($\sim 4.5 - 5 \ \mathrm{eV}$ at $6 \ \mathrm{keV}$), the highest-precision, near-future spin measurements prior to \newathena\ are still forthcoming.

\begin{figure}[H]
\centering 
\includegraphics[width=11.8cm]{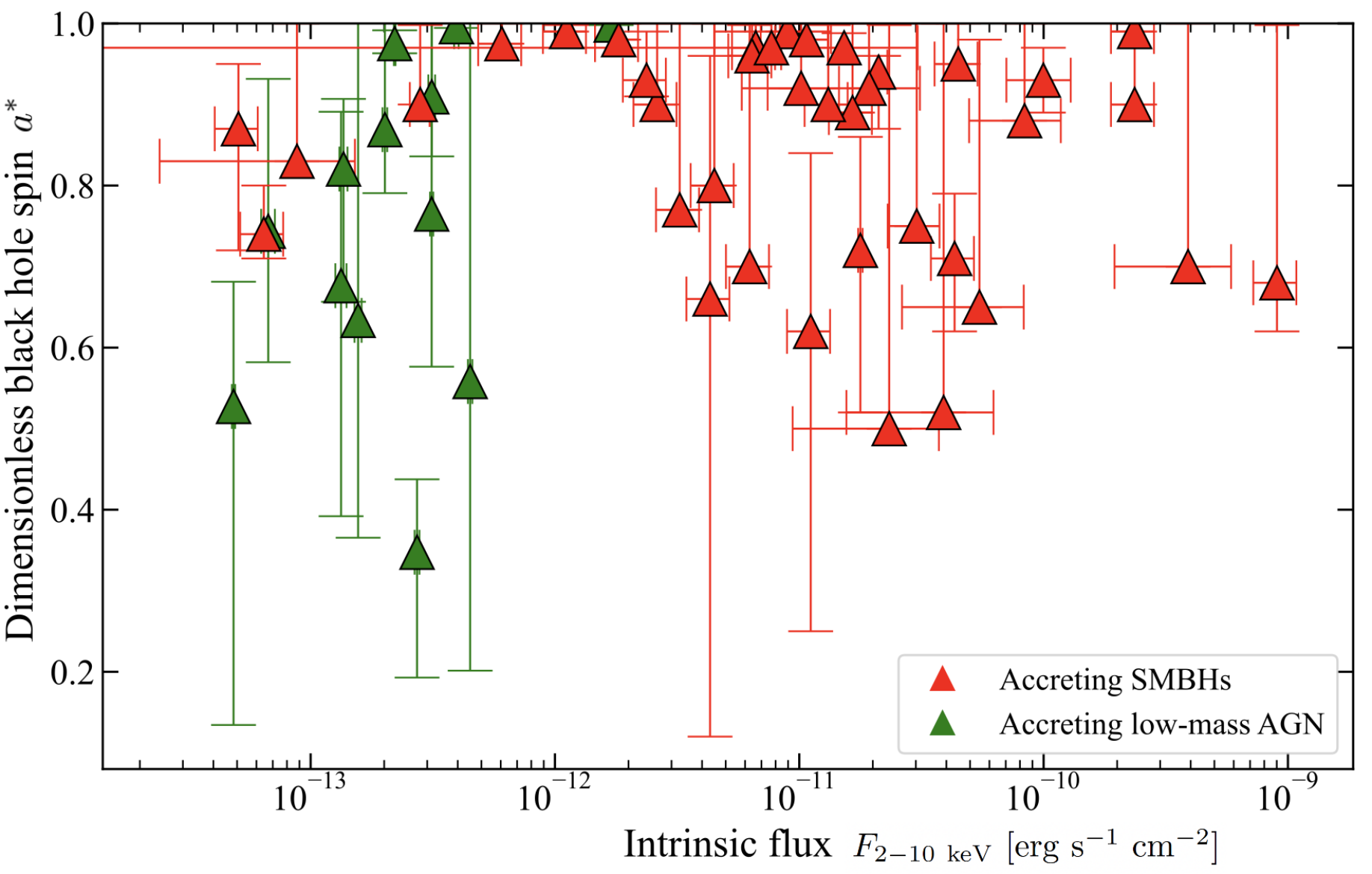}
\caption{\textls[-15]{{Black hole spin} vs. intrinsic (rest-frame) X-ray flux for the full sample of accreting low-mass AGN and SMBHs with reflection-inferred spins. The intrinsic fluxes were estimated using the X-ray luminosity values listed in Table~\ref{tab2} by computing each source's luminosity distance assuming the \textit{Planck} 2018 set of cosmological parameters~\cite{planck-2018}. We estimate the $\sim$1$\sigma$ uncertainties in the flux using those in X-ray luminosity, as follows. Where available, $1\sigma$~statistical uncertainties on the X-ray luminosity from the literature are considered; otherwise, statistical uncertainties of $\pm 20\%$ are considered.} \label{fig02}}
\end{figure}

Beyond the systematic uncertainties inherent to current state‑of‑the‑art reflection{\linebreak} \textls[-15]{\mbox{models---which}, as discussed in Section~\ref{sec3}, generally tend to bias spin estimates toward higher \mbox{values---one} must also take into account spin‑dependent observational selection effects. These were first explored in Ref.~\cite{brenneman-2011} and reviewed comprehensively in \mbox{Ref.~\cite{reynolds-spinreview}---considering} that most existing reflection-based SMBH spin measurements are drawn from broadly flux‑limited samples. Under a given accretion rate $\dot{m}$, high‑spin SMBHs are overrepresented in such samples because their spin‑dependent radiative efficiency $\eta$(\spinpar) makes them intrinsically more luminous (see also Ref.~\cite{piotrowska-2024}). This bias is hinted at in Figure~\ref{fig02}, which shows that---for a given estimated black hole spin---SMBHs with higher intrinsic X‑ray fluxes are more prevalent than their fainter AGN counterparts, albeit with large statistical uncertainties.}

\vspace{-2pt}
\vspace{-8pt}

\begin{table}[H] 
\caption{Ancillary data for the sample of 51 accreting SMBHs (including 13 low-mass AGN studied by Ref.~\cite{mallick-2022}) with X-ray-reflection-inferred black hole spins (where sources appear in the same order as in Table~\ref{tab1}). Where available, each column displays: $L_X$, the intrinsic, absorption-corrected rest-frame 2--10~keV X-ray luminosity; the {Eddington ratio $\lambda = L_\mathrm{bol}/L_\mathrm{Edd}$}; and the redshift, $z$.}
\label{tab2}
\small
\begin{tabularx}{\linewidth}{%
m{3.4cm}<{\centering}
m{3.6cm}<{\centering}
m{3.6cm}<{\centering}
m{1.8cm}<{\centering}
}

\toprule
\textbf{Source} & \boldmath{$L_\mathrm{X,2\text{--}10}~[10^{42} \ \mathrm{erg/s}]$}  & \textbf{\boldmath{Eddington Ratio, $\lambda$}} & \boldmath{$z$} \\
\midrule
H 1821+643 & $3.39 \times 10^3$~\cite{sisk-reynes-2022} & $0.39\pm 0.20$~\cite{fukuchi-2022} & 0.299 \\
Q 2237+305 & $1.30 \times 10^3$~\cite{mark-reynolds-2014} & $\sim$0.01~\cite{mark-reynolds-2014} & 1.695 \\
Fairall 9 & $\sim$240~\cite{leighly-2000} & $\sim$0.15~\cite{lohfink-2012} & 0.047 \\
Ark 120~$^{\dagger}$ & $2.42\times 10^3$~\cite{porquet-2019}& $0.24 \pm 0.08$~\cite{porquet-2019}& 0.033\\
{RXJ 1131–1231}~$^{\dagger}$ & $\sim$100~\cite{reis-2014} & $\sim$0.07~\cite{reis-2014} & 0.658 \\
IRAS 09149-6206~$^{\dagger}$ & $175^{+15}_{-15}$~\cite{walton-2020} & $\sim$0.4~\cite{walton-2020} & 0.057\\
PG 1229+204 & $\sim$25~\cite{zhou-zhao-2010} & $0.002$~\cite{zhou-zhao-2010} & 0.064 \\ 
Swift J2127.4+5654 & $9.6\pm 0.2$~\cite{sanfrutos-2013} & $\sim$$0.14 \pm 0.03$~\cite{marinucci-2014} & 0.015 \\ 
NGC 5506~$^{\dagger\dagger}$ & $8.5 \pm 2.5$~\cite{guainazzi-2010} & $\sim$0.4~\cite{guainazzi-2010} & 0.006 \\ 
Mrk 359 & $\sim$3~\cite{middei-2020} & $\sim$0.08~\cite{middei-2020} & 0.017 \\ 
J0107 & $10^{-1} \times (31.3\pm 0.7)$~\cite{mallick-2022} & ${0.28}^{+0.48}_{-0.19}$~\cite{mallick-2022} & 0.077 \\
J0940 & $10^{-1} \times (37.7\pm 0.8)$~\cite{mallick-2022} & ${0.36}^{+0.59}_{-0.24}$~\cite{mallick-2022} & 0.061 \\
J1357 & $10^{-1} \times (85.3\pm 1.9)$~\cite{mallick-2022} & ${1.0}^{+1.7}_{-0.7}$~\cite{mallick-2022} & 0.106 \\ 
J1541 & $10^{-1} \times (37.5\pm 1.2)$~\cite{mallick-2022} & ${0.35}^{+0.59}_{-0.24}$~\cite{mallick-2022} & 0.068  \\ 
J1140 & $10^{-1} \times (38.2\pm 0.3)$~\cite{mallick-2022} & ${0.45}^{+0.76}_{-0.31}$~\cite{mallick-2022} & 0.081  \\ 
J1347 & $10^{-1} \times (33.0\pm 0.5)$~\cite{mallick-2022} & ${0.47}^{+0.80}_{-0.32}$~\cite{mallick-2022} & 0.064  \\ 
J1434 & $10^{-1} \times (3.0\pm 0.1)$~\cite{mallick-2022} & ${0.04}^{+0.08}_{-0.03}$~\cite{mallick-2022} & 0.028  \\
J1631 & $10^{-1} \times (3.1\pm 0.2)$~\cite{mallick-2022} & ${0.05}^{+0.08}_{-0.03}$~\cite{mallick-2022} & 0.043  \\ 
J1023 & $10^{-1} \times (12.8\pm 0.2)$~\cite{mallick-2022} & ${0.29}^{+0.51}_{-0.19}$~\cite{mallick-2022} & 0.099  \\ 
J1626 & $10^{-1} \times (3.8\pm 0.2)$~\cite{mallick-2022} & ${0.08}^{+0.12}_{-0.05}$~\cite{mallick-2022} & 0.034  \\ 
J0228 & $10^{-1} \times (18.4\pm 0.7)$~\cite{mallick-2022} & ${0.75}^{+1.24}_{-0.50}$~\cite{mallick-2022} & 0.072  \\
POX 52 & $10^{-1} \times (4.8\pm 0.1)$~\cite{mallick-2022} & ${0.15}^{+0.15}_{-0.08}$~\cite{mallick-2022} & 0.021 \\
\midrule
PG 1426+015~$^{\dagger}$ & $126$~\cite{zhou-zhao-2010} & $\sim$0.04~\cite{walton-2025} & 0.087 \\ 
PG 2112+059 & $73 \pm 53$~\cite{saez-2021} & $\sim$0.08~\cite{gallagher-2004} & 0.459 \\ 
PG 0804+761 & $208\pm 20$~\cite{vasudevan-2007} & $\sim$0.4~\cite{pop-2018} & 0.100 \\
{1 H0419–577}~$^{\dagger}$ & $315\pm 70$~\cite{jiang-2018-spins-1h0419} & $\sim$$0.39\pm 0.09$~\cite{jiang-2018-spins-1h0419} & 0.104 \\
Mrk 1501~$^{\dagger}$ & $\sim$140~\cite{schnopper-1978} & $\sim$0.1~\cite{inoue-2007} & 0.089\\
RBS 1124~$^{\dagger}$ & $\sim$600~\cite{minniuti-2010} & $\sim$0.145~\cite{madathil-pottayil-2024} & 0.208 \\
Fairall 51 & $14.2\pm 3.4$~\cite{svoboda-2015} & $\sim$0.025~\cite{svoboda-2015} & 0.014 \\
Mrk 841 & $125 \pm 75$~\cite{vasudevan-2010} & 0.073~\cite{vasudevan-2010} & 0.036 \\
{IRAS 13197–1627}~$^{\dagger}$ & $\sim$240~\cite{minniuti-2010} & $0.05 \pm 0.025$~\cite{vasudevan-2010} & 0.016\\
3C~120 &$120$~\cite{ballantyne-2004} & $\sim$0.77~\cite{ballantyne-2004} & 0.033 \\ 
Mrk 79 & $62.6 \pm 37.4$~\cite{vasudevan-2010} & $0.033 \pm 0.002$~\cite{vasudevan-2010} & 0.033 \\ 
{IRAS 00521–7054}~$^{\dagger}$ & $\sim$40~\cite{walton-2019} & $\approx$1~\cite{walton-2020} & 0.069 \\
NGC 4151~$^{\dagger\dagger}$ & $\sim$5~\cite{keck-2015} & $\sim$0.01--0.1~\cite{bentz-2022} & 0.003 \\ 
1 H0323+342~$^{\dagger}$ & $\sim$25~\cite{rosa-2025} & $\sim$0.18~\cite{rosa-2025} & 0.061\\
{ESO 033–G002}~$^{\dagger}$ & $\sim$5~\cite{walton-2021} & $\sim$0.02~\cite{walton-2021} & 0.018\\
NGC 3783 & $19.9\pm 8.1$~\cite{vasudevan-2010} & $0.06\pm 0.01$~\cite{brenneman-2011} & 0.010 \\ 
Mrk 110 & $\sim$50~\cite{porquet-mrk110} & $\sim$0.1~\cite{porquet-mrk110} & 0.035 \\
Mrk 335 & $16 \pm 8$~\cite{leek-2016} & 0.005--0.04~\cite{sarma-2015} & 0.027 \\
PG 1535+547~$^{\dagger}$ & $\sim$4~\cite{madathil-pottayil-2026} & $0.315$~\cite{madathil-pottayil-2026} & 0.038 \\ 
{ESO 362–G18}~$^{\dagger\dagger}$ & <5.1~\cite{agis-gonzalez-2014} & $\sim$0.02~\cite{agis-gonzalez-2014} & 0.012 \\
Tons 180 & $\sim$18~\cite{matzeu-2020} & >0.55~\cite{matzeu-2020} & 0.062 \\
IRAS 13224–3809~$^{\dagger}$ & $6.82$~\cite{jiang-2018-spins-iras13224} & $0.32\pm 0.05$~\cite{jiang-2018-spins-iras13224} & 0.066\\
{1 H0707–495}~$^{\dagger}$ & NA & $\approx$1~\cite{done-2016} & 0.041\\
{MCG–06-30-15}~* & $8.3\pm 4.3$~\cite{vasudevan-2010} & $\sim$0.08~\cite{brenneman-2025} &  0.008 \\
Mrk 1044~$^{\dagger}$ & $8.6\pm 0.8$~\cite{barua-2023} & $0.34\pm 0.09$~\cite{barua-2023} & 0.106\\
Ark 564 & $\sim$20~\cite{turner-2001} & NA & 0.025\\
NGC 1365 & $1.3 \pm 1.3$~\cite{vasudevan-2010} & $0.03\pm 0.01$~\cite{vasudevan-2010} & 0.006 \\
Mrk 766~$^{\dagger}$ & $7.8 \pm 4.8$~\cite{vasudevan-2010} & $0.04 \pm 0.02$~\cite{vasudevan-2010} & 0.013 \\
J1559 & $10^{-1} \times (40.1\pm 0.1)$~\cite{mallick-2022} & ${0.38}^{+0.63}_{-0.25}$~\cite{mallick-2022} & 0.031\\ 
\bottomrule
\end{tabularx}
\noindent{\footnotesize{\textls[-9]{
The following notation highlights accreting SMBHs whose spins (as listed in Table~\ref{tab1}) were inferred from either broadband multi-epoch \xmm+\nustar\ data~$^\dagger$, simultaneous \xmm+\nustar+\xrism\ coverage~*, other broadband datasets~$^{\dagger\dagger}$, e.g., \suzaku/XIS+\xmm, or no multi-epoch broadband coverage (no~symbol). `NA' indicates parameters of interest which were either not available in the literature or which could not be estimated from the literature at the time of writing.}}}
\end{table}

The current observed sample is also heterogeneous in X-ray luminosity, Eddington ratio, spectral type, and redshift. Where available, these quantities are listed in Table~\ref{tab2}.

\begin{figure}[H]
\includegraphics[width=13.8cm]{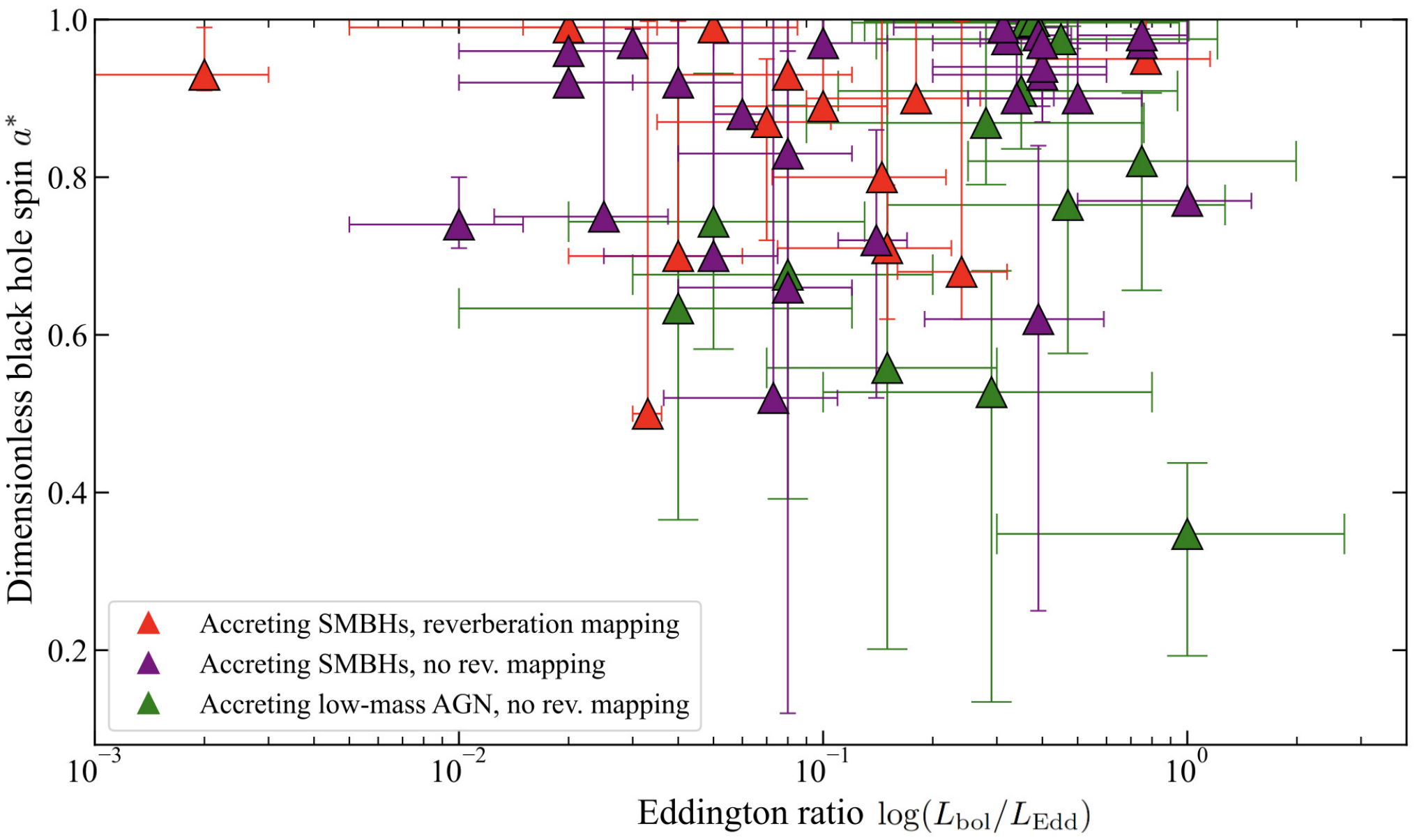}
\caption{Black hole spin versus Eddington ratio---defined as the ratio of the bolometric luminosity to the Eddington luminosity---for the full sample of accreting low-mass AGN (green) and other SMBHs \mbox{(red + purple)} with reflection-inferred spins.~The latter sample of 38 SMBHs is split into two distinct sub-samples: one where the black hole mass was estimated from optical reverberation mapping (red), and another where the mass was inferred by other methods (purple). The black hole masses of the 13 low-mass AGN in Ref.~\cite{mallick-2022}~were inferred using methods other than reverberation mapping. Where available, $1\sigma$ statistical uncertainties {in} the Eddington ratio from the literature are considered; otherwise, statistical uncertainties of $\pm 50\%$ are shown. \label{fig2}}
\end{figure}

Figure~\ref{fig2} shows no obvious correlation between spin and Eddington ratio for the full sample. This result {does not provide clear observational support for} models claiming an Eddington-ratio dependent equilibrium spin (e.g.,~\cite{ricarte-2023}). However, SMBHs need not be at the equilibrium spin for a given Eddington ratio if the Eddington ratio fluctuates on short timescales. {Ref.~\cite{lowell-2025} found a low universal equilibrium spin of $a^*\approx0.3$ for luminous, thin magnetically arrested disks (MADs)---suggesting that the data in Figure~\ref{fig02} could be explained if most accretion does \textit{not} proceed in such a highly magnetized fashion}. 
 
\begin{figure}[H]
\includegraphics[width=13.8cm]{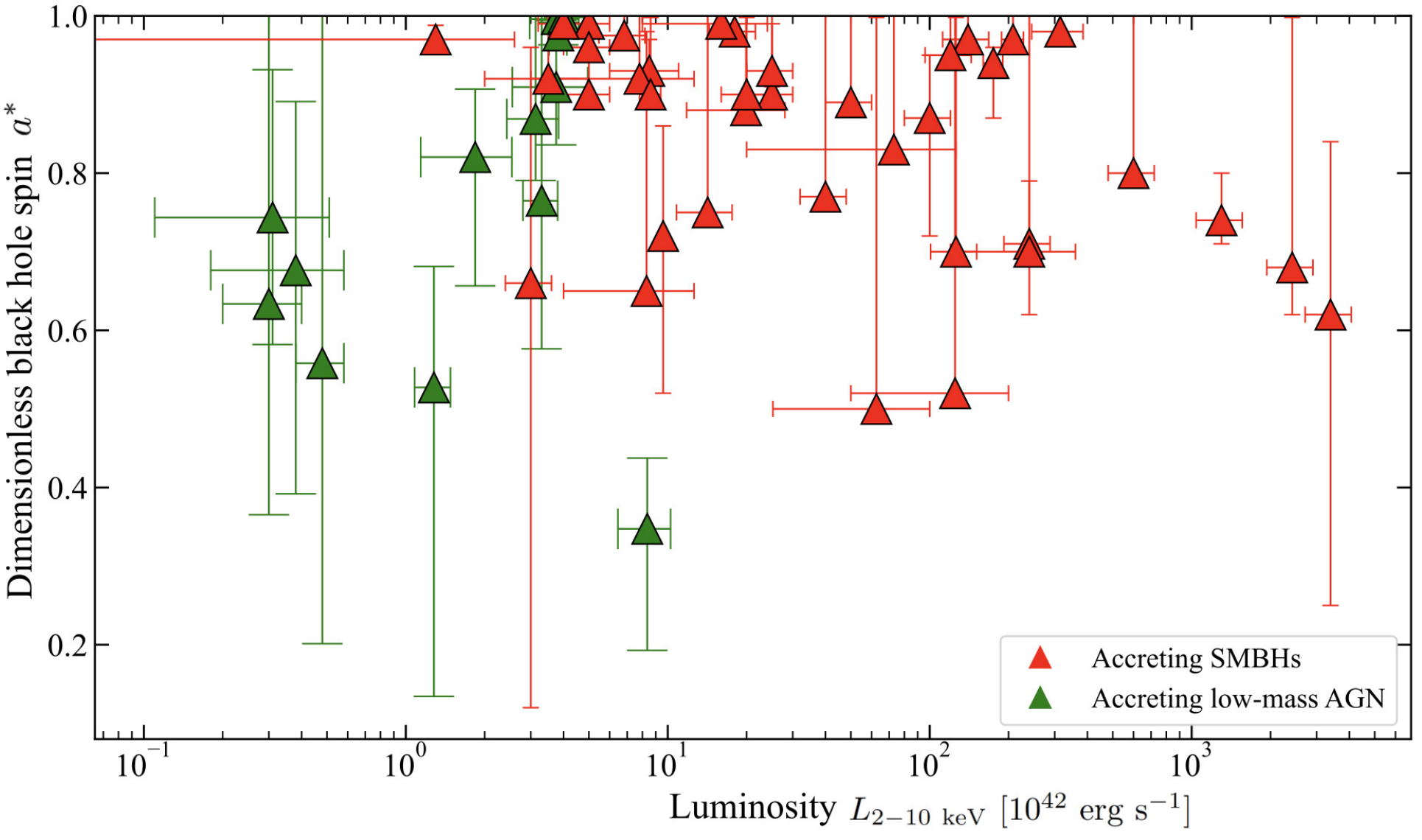}
\caption{Black hole spin versus intrinsic (absorption-corrected) X-ray luminosity (2--10~keV) for the full sample of accreting low-mass AGN and SMBHs with reflection-inferred spins. Where available, $1\sigma$~statistical uncertainties on the X-ray luminosity from the literature are considered; otherwise, statistical uncertainties of $\pm 20\%$ are considered. \label{fig3}}
\end{figure}

\textls[-9]{Future studies using large, homogeneous samples will need to account for the known correlation between the bolometric correction and Eddington{-scaled accretion rate} (particularly if the Eddington ratio is used to estimate the bolometric luminosity from the 2--10~keV luminosity), since neglecting this correlation could bias Eddington ratio estimates~\cite{vasudevan-2007,vasudevan-2016-sample}. We note that most works in the literature adopt bolometric corrections from the observed 5100~\r{A} luminosity. Each of these approaches carries its own caveats: the AGN contribution to $L_{5100}$ is limited by the host‑galaxy subtraction method \citep{runnoe-2012}, although the {5100 \r{A}} bolometric corrections are generally preferred because they exhibit smaller intrinsic scatter than X‑ray bolometric corrections. In parallel, the estimate of 2--10 keV bolometric correction is itself correlated with the Eddington fraction, which---if unaccounted for---can lead to systematically overestimated values of $\lambda_\mathrm{Edd}$ \citep{vasudevan-2009,vasudevan-2016-sample}. Figure~\ref{fig3} also shows no obvious correlation between spin and the 2--10 keV luminosity in the sample.}

\textls[-8]{We note that uncertainties in black hole mass and Eddington ratio estimates---arising from the use of different methods such as reverberation mapping, $H\beta$ or $C$\texttt{IV} line width diagnostics, and other approaches listed in Table~\ref{tab1}---also contribute to the overall scatter in the observed trends. These methodological differences represent an additional, non‑negligible component of the uncertainties discussed in the following section, which explores the challenges and limitations associated with current spin measurements and their impact on spin demographic studies.}

\section{Future Prospects: A Decisive Test of Observed Mass--Spin Trends with~\newathena}
\label{sec3}

\textls[-8]{The current observed SMBH mass--spin sample, while invaluable, remains dominated by lower spin limits and is too heterogeneous to enable decisive tests of SMBH growth scenarios. This limitation has been demonstrated explicitly by Ref.~\cite{piotrowska-2024}, who showed that the {collection} of reflection‑based spin measurements at the time lacked the statistical power and uniformity required to distinguish between the distinct mass--spin trends predicted for coherent‑accretion‑dominated versus coherent‑accretion+merger‑dominated SMBH growth from the \textsc{Horizon-AGN} cosmological simulation. Ref.~\cite{piotrowska-2024} further showed that strategically sampling the SMBH mass–spin plane with the proposed {\em High‑Energy X‑ray Probe} (\textit{HEX‑P})~\cite{madsen-hexp-2024}---a mission concept with \nustar-like high‑energy coverage but a substantially larger collecting area---with three mass bins would enable population‑level spin measurements. Their approach drew a direct connection to the distinct accretion versus accretion+merger driven growth channels in \textsc{HorizonAGN} by enabling observational tests of their distinct trends in the mass--spin plane, subject to specific requirements in sample size, mass range, and redshift coverage (outlined in Section 7 of Ref.~\cite{piotrowska-2024}).}

In addition to sample size limitations, the assessment of systematic uncertainties in spin measurements is equally crucial~\cite{barret-2019}. These systematics fall broadly into two categories: (i) modeling systematics, arising from both spectral degeneracies and assumptions made by relativistic reflection models, and (ii) instrumental systematics, arising from calibration uncertainties. Recent work has also shown that within the first category, spectral model degeneracies---particularly those involving warm absorbers, ultra‑fast outflows, and complex soft X‑ray structure---can mimic or obscure relativistically smeared reflection features. Ref.~\cite{parker-degeneracies} demonstrated this explicitly in the context of \newathena\ and also showed that these degeneracies are dramatically reduced when broadband coverage is included to fully probe the Compton hump and characterize the high-energy cutoff. In parallel, a comprehensive analysis that jointly incorporates the impact of different modeling assumptions on the reflection-based spin---such as alternative coronal geometries beyond the lamppost, the use of razor‑thin versus razor‑thick disk prescriptions, or considering reflected emission within the innermost stable circular orbit---has yet to be completed, although several recent studies have begun to quantify how such choices can affect the recovered spin for a given underlying spin (e.g., Refs.~\cite{taylor-2018,delilah-2025,nekrasov-2025}).

Within this first category, the role of disk density is also increasingly recognized as a key modeling uncertainty. Standard implementations of \textsc{Relxill}~assume a fixed mid‑plane density of $n_\mathrm{e} = 10^{15} \ \mathrm{cm}^{-3}$, which directly sets the ionization parameter and is known to be an oversimplification. High‑density reflection models relax this assumption by allowing the density to vary, and in several cases alleviate the need for the super‑solar inner disk iron abundances often inferred with fixed‑density models. However, the interplay between density, ionization, and the inner disk metallicity remains poorly understood, and a systematic assessment of how high‑density models impact recovered spin values has not yet been performed. In parallel, a further source of uncertainty arises from the poorly constrained inclination distribution of current reflection‑based spin measurements: most studies report only approximate inclination values, preventing a proper assessment of correlations between inclination, spin, and other spectral parameters. Very high inferred inclinations can partially mimic the blueshifted wing of the broadened $\mathrm{Fe} \ \mathrm{K}\alpha$ line, introducing a degeneracy that cannot yet be quantified robustly in the absence of full posterior information for most published analyses.

\textls[-13]{Within the second category, instrumental calibration remains an important consideration. Ref.~\cite{sisk-reynes-2023} demonstrated that machine‑learning approaches such as convolutional neural networks (when trained to distinguish astrophysical spectral features from detector‑calibration artifacts in synthetic \newathena\ spectra) offer a promising pathway towards mitigating these systematics. These developments highlight that progress in SMBH spin studies requires not only larger samples but also improved tools for disentangling physical and instrumental effects. We highlight that careful consideration of systematic uncertainties will be especially critical for \newathena, whose dramatically reduced Poisson uncertainties---enabled by its much larger collecting area---will shift the limiting factor in reflection-based spin estimates from statistical noise to these systematic effects.  }

The upcoming \newathena\ X‑ray observatory is poised to deliver the uniform dataset of reflection-based spin estimates identified as necessary by Ref.~\cite{piotrowska-2024}, while with a mission profile distinct from \hexp. {The science requirements of \newathena\ anticipate a survey of at least 50 nearby SMBHs with $\leq$10\% statistical precision per spin estimate, delivering the first homogeneous}, high‑precision spin catalog for the local, intermediate‑mass AGN population, providing a well‑characterized anchor sample for confirming tentative mass–spin predictions. As highlighted by Section 7 of Ref.~\cite{piotrowska-2024}, a sample of 50 accreting SMBHs may neither overcome the high-spin bias reflected in the current sample, nor uniformly populate the high-mass SMBH end ($M_\mathrm{BH} > 10^9\,M_\odot$) in the local universe, where theoretical models of accretion-driven vs. accretion + merger-driven growth differ more significantly. Noting that at $z\sim 0$ only a handful of such massive, X-ray luminous SMBHs exist, the forthcoming \newathena\ spin survey of at least 50 nearby AGN will be dominated by SMBHs with intermediate masses. For this reason, the current heterogeneous sample of 51 reflection-based spin estimates will remain essential for spanning the full 10$^{6}$--10$^{10}\,M_\odot$ mass range required to statistically discriminate between competing mass--spin trends in cosmological models and SAMs. However, we note that many of the existing estimates will need to be revisited---and reobserved with \newathena\ if \xrism\ follow-up is not available at the time of \newathena's launch---before such estimates can be used along with \newathena's new survey. This is particularly pressing for current spin estimates driven by the reflection interpretation of the soft excess, which has led to high \spinpar\ estimates in several Narrow-Line Seyfert 1 galaxies in the current sample. In such cases, the soft excess itself can act as the dominant driver of the spin constraint: in the absence of a pronounced broadened $\mathrm{Fe}~\mathrm{K}\alpha$ line and Compton hump, the forest of relativistically blurred soft X-ray features predicted by reflection models forces the fit toward high spin, effectively yielding high-spin lower limits that may \textit{not} reflect the true underlying spin.

\textls[-5]{Probing spin evolution across cosmic time will require extending beyond the local sample, since a nearby survey alone cannot map redshift‑dependent trends. The value of the \newathena\ survey will instead lie in providing a well‑characterized anchor sample with unprecedented sensitivity due to its improved collecting area and unprecedented spectral resolution compared to current X-ray missions. We highlight that for distant SMBHs up to $z\sim 1.5$, the rest‑frame Fe K band is redshifted into \newathena’s 0.1--12~keV sensitivity window, enabling spin measurements for bright sources at moderate redshift without relying on strong gravitational lensing. This redshift effect opens a complementary pathway for extending spin studies beyond the local universe, although the achievable precision will depend on source brightness and exposure time.}

\subsection*{A Statistical Framework to Probe SMBH Mass--Spin Trends with NewAthena}
\textls[-5]{Hierarchical Bayesian inference offers a particularly powerful and timely tool to extract the underlying spin distribution from both current and future observed mass--spin datasets. This framework has already proven transformative in gravitational‑wave astronomy, where it is routinely used by the \textit{LVK} Collaboration to infer the underlying spin distributions of merging stellar‑mass black holes from gravitational-wave signals~\cite{lvk-collaboration}. Applying analogous techniques to \newathena’s spin catalog will allow the X-ray community to start addressing if the local SMBH population reflects a mixture of coherent and {chaotic} accretion, SMBH mergers, and jet‑induced spin‑down, to quantify the relative importance of these channels at low redshift. These techniques will enable incorporating systematic uncertainties due to spectral model degeneracies, modeling assumptions, and detector calibration, to account for spin‑dependent radiative‑efficiency selection effects in flux‑limited samples, and to enable principled comparisons between physical models of SMBH growth. Such hierarchical population analyses have not yet been attempted for SMBH spins inferred from X-ray reflection, and the first exploratory tests in the context of \newathena\ are underway \cite{newathena-inprep}. Preliminary exploration using a small set of \mbox{hyperparameters---capturing} a merger-driven population as a Gaussian about a mean with moderate \spinpar\ and a truncated power-law function underlying a SMBH population growing primarily via coherent accretion---indicate that this two-component description fails to reproduce the observed mass--spin distribution, illustrating the need for more complex and physically motivated models. The current heterogeneous mass--spin sample of 51 SMBHs---though not yet statistically \mbox{decisive---provides} an ideal laboratory for developing such techniques prior to their application across the \newathena~dataset.}

In addition to the spin constraints provided by \newathena, achieving this goal will require SMBH mass estimates of the highest possible precision, drawing on complementary facilities such as the \textit{Nancy Grace Roman Space Telescope} to provide robust black hole mass estimates. Moreover, the observational requirements implemented by Ref.~\cite{piotrowska-2024} for the \hexp\ spin survey will need to be revisited in the context of \newathena, since its spectral capabilities, survey strategy, and redshift reach differ from those assumed in the \hexp\ forecasts of Ref.~\cite{piotrowska-2024}. In particular, unlike \hexp, \newathena\ will \textit{not} have the high-energy coverage needed to fully characterize the Compton hump or robustly constrain the high-energy cutoff---essential for tight reflection-based spin estimates and to break spectral model degeneracies. Thus, assessing the impact of this limitation and determining how it could affect the distinction between competing mass--spin trends will be a central component of \newathena's SMBH spin survey. 

Together, \newathena\ and hierarchical Bayesian population modeling promise to transform SMBH spin studies with observed SMBH mass--spin distributions. Whilst this combination cannot realistically capture the full SMBH evolution in mass and redshift, it will provide the most robust and controlled benchmark of observed mass--spin trends in the local universe, transforming these observed correlations into a quantitative test of competing growth scenarios.

\section{Conclusions}
\label{sec4}
\begin{itemize}
    \item {We have compiled an updated and comprehensive census of SMBH spin measurements obtained via relativistic X‑ray reflection spectroscopy, consolidating a heterogeneous literature into a single resource: the Github repository \href{https://github.com/joanna-pk/xray-reflection-spin-repository}{https://github.com/joanna-pk/xray-reflection-spin-repository} (accessed on 10 May 2026).
    We have highlighted this method’s unique ability to probe the angular momentum of SMBHs embedded in optically thin, geometrically thick accretion flows.}
    \item {SMBH spin demographics have the potential to offer a powerful probe of recent black hole growth, but the present mass--spin sample remains too heterogeneous to support decisive population-level inferences. Large statistical uncertainties, inconsistent data quality, differing modeling assumptions, limited broadband coverage (with $\sim$50\% of current estimates based on broadband X-ray spectra covering both the Fe K band and the Compton hump), and the fact that only 22/51 spins are well‑constrained contribute to substantial scatter and preclude formal correlation analyses.}
    \item {The current sample also suffers from structural limitations, including heteroscedastic spin uncertainties, under-representation of high-mass SMBHs (>10$^8\,M_\odot$), and methodological diversity in mass and Eddington ratio estimates---which collectively hinder efforts to extract robust trends in the mass--spin plane or to discriminate between competing SMBH growth scenarios at $z = 0$.}
     \item {\newathena’s anticipated survey of $\geq$50 SMBHs (with <10\% statistical precision in spin recovery and sensitivity to high‑redshift AGN whose Fe K band is redshifted into the X‑IFU bandpass) will provide the first opportunity to populate the mass–spin plane in a statistically meaningful way. This will carve the pathway for decisive tests of SMBH growth models and, for the first time, allow spin measurements of luminous AGN at cosmological distances without relying on the strong lensing flux magnification.}
     \item {Even with \newathena’s transformative capabilities, robust inference of physical trends in the data will require methods that can incorporate instrumental systematics, spectral degeneracies, and model‑dependent uncertainties. Hierarchical Bayesian inference offers a promising framework for jointly modeling these effects and extracting reliable population‑level constraints on SMBH spin evolution from future homogeneous and high-quality datasets}.
\end{itemize}

\vspace{3pt} 


\authorcontributions{J.M.S.-R. led the contextualization, data preparation, and writing of this manuscript. Conceptualization: J.M.S.-R., C.S.R, J.H.M, and D.J.W. Validation: J.M.S.-R. and J.M.P. Writing—review: J.M.S.-R., D.J.W, J.M.S and A.R. All authors have read and agreed to the published version of the manuscript.}

\dataavailability{All the data presented are public and available at \url{https://github.com/joanna-pk/xray-reflection-spin-repository} (accessed on 10 May 2026).} 

\acknowledgments{\textls[-29]{J.M.S.-R. acknowledges support from a NASA ADAP Program Grant 80NSSC24K0617}. D.J.W. acknowledges support from the Science and Technology Facilities Countil (STFC; grant code ST/Y001060/1). {J.F.S acknowledges support from NASA Grant NAS8-03060}. J.M.S.-R. thanks Labani Mallick for sharing a digitized version of the data presented in Ref.~\cite{mallick-2022} and {Laura Brenneman, Matteo Guainazzi, and Daniel Schwartz} for comments on this~manuscript.~We are also grateful to the three referees for their insightful comments, which improved the quality of this manuscript. We are grateful to the organizers of the `Taking Spin Measurements for a Spin:
Recent Progress on Black Hole Spin Measurements Across the Electromagnetic and Gravitational Spectra' Workshop at Wake Forest University in September 2025 for triggering insightful discussions that led to the preparation of this manuscript. J.M.S.-R. is grateful to Alejandro C\'ardenas-Avendaño, Delilah Gates, and George Wong for their technical assistance through the preparation of this manuscript.}

\conflictsofinterest{The authors declare no conflicts of interest.} 

\begin{adjustwidth}{-\extralength}{0cm}
\setenotez{backref=true,list-name={Note}}
\printendnotes[custom] 

\reftitle{References}

\PublishersNote{}
\end{adjustwidth}
\end{document}